\documentstyle[12pt]{article}
\newcommand{\draft}{
        \renewcommand{\baselinestretch}{1.0}%
        \small\normalsize%
}

\draft
\begin{document}
\title{\bf Effects of the background radiation on radio pulsar and 
supernova remnant searches and the birth rates of these objects} 
\author{A\c{s}k\i n Ankay$\sp1$    
\thanks{e-mail:askin@gursey.gov.tr},
Oktay H. Guseinov$\sp{1,2}$
\thanks{e-mail:huseyin@gursey.gov.tr},
Sevin\c{c} O. Tagieva$\sp3$
\thanks{email:physic@lan.ab.az}, \\ \\
{$\sp1$T\"{U}B\.{I}TAK Feza G\"{u}rsey Institute,} \\
{81220 \c{C}engelk\"{o}y, \.{I}stanbul, Turkey} \\
{$\sp2$Akdeniz University, Department of Physics,} \\
{Antalya, Turkey} \\
{$\sp3$Academy of Science, Physics Institute, Baku 370143,} \\
{Azerbaijan Republic}} 

\date{}
\maketitle
\begin{abstract}
\noindent
In different directions of the Galaxy the Galactic background radio 
radiation and radiation of complex star formation regions which include 
large number of OB associations have different influences on radio pulsar 
(PSR) and supernova remnant (SNR) searches. In this work we analyse the 
effects of these background radiations on the observations of PSRs at 
1400 MHz and SNRs at 1000 MHz. In the interval l=0$^o$$\pm$60$^o$ the 
PSRs with flux F$_{1400}$$>$0.2 mJy and the SNRs with surface brightness 
$\Sigma$$>$10$^{-21}$ Wm$^{-2}$Hz$^{-1}$sr$^{-1}$ are observable for all 
values of l and b. All the SNRs with $\Sigma$$>$3$\times$10$^{-22}$ 
Wm$^{-2}$Hz$^{-1}$sr$^{-1}$ can be observed in the interval 
60$^o$$<$l$<$300$^o$. We have examined samples of PSRs and SNRs to 
estimate the birth rates of these objects in the region up to 3.2 kpc 
from the Sun and also in the Galaxy. The birth rate of PSRs is about one 
in 200 years and the birth rate of SNRs is about one in 65 years in our 
galaxy.    
\end{abstract}


\clearpage
\parindent=0.2in

\section{Introduction}
Although the problem of birth rates of radio-pulsars (PSRs) and 
supernova remnants (SNRs) has been discussed for many years, it is still 
an open question. The main difficulties in solving this problem are the 
selection effects in observations. It is not well known under which 
circumstances single PSRs are born under a supernova (SN) 
explosion. Actually, PSRs may be born under SN explosions of types Ib, 
Ic and II (here we do not consider the accretion induced collapse).
SNRs are formed as a result of SN 
explosions with energies 10$^{49}$-10$^{51}$ erg, and in some cases, even 
with energies several times smaller than
10$^{49}$ erg (e.g. Crab, Sollerman et al. 2000) and with energies
$>$10$^{51}$ erg (e.g. Cas A, Vink et al. 1998; Wright et al. 1999). 

According to Lorimer et al. (1993) PSRs are formed once every 150 years in 
the Galaxy 
and the lower limit for the mass of the stars which form PSRs at the end
of their evolution is about 5 M$_{\odot}$. By examining the historical
SNRs, Strom (1994) found that a SN explosion occurs every 6 years in the 
Galaxy and the lower limit for the mass of the progenitors of these SNRs 
is also about 5 M$_{\odot}$. There is an unlikely large difference between 
the birth rate of PSRs given by Lorimer et al. (1993) and the formation 
rate of SNe given by Strom (1994). Does the formation of PSRs 
predominantly depend on some other parameters because of the lower limit 
for the progenitor mass being the same in both cases?

It is known that in Sb-type galaxies rate of SN
explosion is similar to the SN rate in our galaxy, $\sim$1-2 in 100
years on average, because our galaxy also is Sb-type.
(van den Bergh \& Tammann 1991). Recent statistical investigations of SN 
rate in Sb-type galaxies show that the rate of SN Ia is 0.4$\pm$0.2 in 100 
years and the rate of SN II together with SN Ib and SN Ic is about 
1.5$\pm$1.0 per century in Sb-type galaxies and so in our 
galaxy (Capellaro et al. 1999; Capellaro \& Turatto 2001).

In our galaxy no evidence was found of a SN explosion in the last 300
years. The results of optical, radio and X-ray observations of the
region up to 5 kpc around the Sun showed that there is a very small
probability to find a neutron star (NS) or a SNR with such a small age. 

Kaspi et al. (1999) and Kaspi \& Helfand (2002) give a list of the 
youngest ($\tau$$\le$2.44$\times$10$^4$ yrs) 17 PSRs. Ten of these PSRs 
are genetically connected to SNRs. 
The opposite of this is not true, i.e. in most of the SNRs with such ages 
no PSR has been found. Searching for dim point X-ray sources in nearby 
SNRs is essential to solve this important problem, because after 
finding point X-ray sources in SNRs these point sources can precisely be 
examined in the radio band.

Assuming the lower limit for the mass of the progenitors, which end their
evolution with SN explosion, to be 5 or 8 M$_{\odot}$ leads to a 
difference of a factor of 3 in the formation rate of SNRs, if
we use the initial mass function (IMF) of Blaha \& Humphreys (1989). 
Even for different galaxies and star formation regions (SFRs) we can use a 
simple IMF with a value of power about 2.3--3 (Schaerer 2002). A lower 
limit for the mass of progenitors of SN about 7-8 M$_{\odot}$ is in 
accordance with a rate of one SN in 65 years. It is necessary also to 
note that the SFRs in our galaxy are not symmetrically located and also 
the star formation rates vary from one region to another. 

\section{Effects of the background radiation on SNR and PSR searches}
\subsection{Effects of the background radiation on each PSR and each SNR
separately}
It is known that the background radio radiation increases when the line of
sight becomes closer to the Galactic center direction and the Galactic 
plane. Distribution of the temperature which characterizes the background
radiation at 400 MHz is known ('Physics of Cosmos' 1986). When we 
compare the intensity (temperature) of the background radiation with the
structure of the Galactic arms (Georgelin \& Georgelin 1976; Paladini et 
al. 2003), the effect of giant HII regions located in SFRs is seen. 

As known, the source of the
background radiation in the radio band is electron gas 
and number density of electrons increases under the approach to
the Galactic center. Mainly because of this reason, as the direction
approaches to the Galactic longitude l=0$^o$ and the Galactic 
latitude b=0$^o$, intensity of this radiation increases. 
Naturally, the background radiation is most effective on the 
PSRs with low flux values and on the SNRs with low surface brightness 
($\Sigma$) values.

In the Galaxy, there are HII region complexes (which include several OB 
associations) with
large sizes and high surface brightness values. Some of these complexes  
are located close to
the Sun and in the directions far away from the Galactic center direction. 
Such HII region complexes can have considerable  
contribution to the background radio radiation and they can change
the smoothness of distribution of the background radiation at small
angular sizes. This is clearly seen, for example, for the  
region where Vela is located in (262$^o$ $<$ l $<$ 268$^o$). Below, we
will discuss the effect of the background radiation and the effects of
different SFRs (which include many O-type stars) on the PSR search at
1400 MHz and on the SNR search at 1000 MHz. Please note that
the searches for PSRs at 400 MHz have high sensitivity only in the Arecibo
window (40$^o$ $<$ l $<$ 65$^o$, $\mid$b$\mid$$\le$2.5$^o$, Hulse \&
Taylor 1974, 1975). It must also be noted that PSRs have steep spectrum in
general, so that, effect of the background radiation is not so important 
on radiation of PSRs at 400 MHz.

We have examined the effect of Galactic background radiation on the
observed SNRs by considering the l and b values. SNRs
G3.8+0.3 and G354.8-0.8 are the dimmest among the SNRs which are the
closest to the Galactic center direction (in the range l=0$^o$$\pm$10$^o$ 
and $\mid$b$\mid$$<$2$^o$). $\Sigma$ values of these 2 SNRs
are, respectively, 1.86$\times$10$^{-21}$ Wm$^{-2}$Hz$^{-1}$ster$^{-1}$   
and 1.17$\times$10$^{-21}$ Wm$^{-2}$Hz$^{-1}$ster$^{-1}$. The SNRs
G6.4+4.0 and G358.0+3.8 (which have a bit larger $\mid$b$\mid$ values) 
have $\Sigma$ values 2.04$\times$10$^{-22}$
Wm$^{-2}$Hz$^{-1}$ster$^{-1}$ and 1.56$\times$10$^{-22}$  
Wm$^{-2}$Hz$^{-1}$ster$^{-1}$, respectively. It is possible to 
observe such low-$\Sigma$ SNRs with $\mid$l$\mid$$>$60$^o$--70$^o$ (i.e. 
far away from the Galactic center) and even with $\mid$b$\mid$$<$2$^o$. 
Among the observed SNRs only 2 of
them (G156.2+5.7, which is not shown in Figure 1, and G182.4+4.3) have 
$\Sigma$ $<$ 10$^{-22}$ Wm$^{-2}$Hz$^{-1}$ster$^{-1}$ (Green 2001). So, 
the effect of the background radiation on the SNR search in the Galactic 
anticenter directions can surely be neglected for the SNRs with $\Sigma$ 
$\ge$ 3$\times$10$^{-22}$ Wm$^{-2}$Hz$^{-1}$ster$^{-1}$ (see Figures 1 
and 2). On the other hand, only 21\% of all the SNRs given in Green 
(2001) have $\Sigma$ $<$ 
10$^{-21}$ Wm$^{-2}$Hz$^{-1}$ster$^{-1}$. For the SNRs in the anti-center 
directions, even if the $\Sigma$ values are small, the flux values 
(F$\sim$$\Sigma$$\times$$\theta$$^2$, where $\theta$ is the angular 
diameter of the SNR) can be larger compared to the flux values of the 
SNRs in the Galactic central directions in most of the cases (see Figure 
3), because the SNRs in the anticenter directions have, in general, 
smaller distances and larger sizes. 

Among the known PSRs in the interval l=0$^o$$\pm$10$^o$ and 
$\mid$b$\mid$$<$2$^o$, PSR J1728-3733 (l = 
350$^o$.8, b = $-$1$^o$.66) has the lowest flux at 1400
MHz: F$_{1400}$ = 0.19 mJy. Other low-flux PSRs are PSR J1804-2228 (l =
7$^o$.72, b = $-$0$^o$.4) with F$_{1400}$ = 0.2 mJy and both PSR
1736-3511 (l = 353$^o$.6, b = $-$1$^o$.6) and PSR J1751-2516 (l =
3$^o$.85, b = 0$^o$.69) with F$_{1400}$ = 0.22 mJy. So, the background   
radiation practically can not hide PSRs in the 
surveys of the last $\sim$10 years if F$_{1400}$ $>$ 0.2
mJy (similar to the case of SNRs with $\Sigma$$>$10$^{-21}$
Wm$^{-2}$Hz$^{-1}$ster$^{-1}$) (Figures 4,5). It is 
necessary to note that the observations of PSRs also depend on the pulse 
period, the dispersion measure and also the observational instruments.
Here it is easier and more reliable 
to make statistical investigations because most of the PSRs were observed 
with the same telescope. Below, we will examine the influence of the 
background radiation and the influence of nearby HII regions on PSR and 
SNR searches.
\subsection{Effect of the background radiation on the samples of PSRs and
SNRs}  
We can assume that, the SNR search in the Galaxy has
been made with roughly the same sensitivity, but not necessarily with the 
same precision, in all directions.
In Figure 3, flux values (at 1000 MHz) of the SNRs
(Green 2001) with respect to Galactic longitude
for the SNRs with $\mid$b$\mid$ $<$ 5$^o$ are represented. As seen in this
figure, the SNRs were searched down to the same flux value in all 
directions, but, since SNRs are extended objects, the SNRs with larger 
angular sizes are more easily observed in the Galactic anticenter 
directions. Observing SNRs depends significantly on their $\Sigma$ values 
as well as their fluxes. 

The distribution of the SNRs in different longitude intervals with 
respect to $\Sigma$ show that the longitude
interval which is the most affected by the background radiation is
l=0$^o$$\pm$40$^o$ (Figure 2). How this effect decreases as the  
line of sight recedes from the Galactic center direction is also
clearly seen. As mentioned above, among the SNRs with 
$\mid$b$\mid$$<$5$^o$ the lowest $\Sigma$ value belongs to SNR  
G182.4+4.3. As seen from Figure 1, in the Galactic central directions, 
except l$\cong$0$\pm$40$^o$, almost all the SNRs with 
$\Sigma$$>$3$\times$10$^{-22}$ Wm$^{-2}$Hz$^{-1}$ster$^{-1}$ are
observable. The background radiation is strong in the  
regions l $\cong$ 10$^o$--30$^o$ and l $\cong$ 330$^o$--340$^o$, and the
number of the HII regions in these intervals is large ('Physics of 
Cosmos' 1986; Georgelin \& Georgelin 1976; Paladini et al. 2003). Since 
this is   
related to the number of massive stars being large in these intervals, this
fact shows itself in Figure 1. In such regions the formation rates of SNRs
and PSRs must be high. The result of this is not clearly seen in Figures
1--3, but it can be seen in Figures 4 and 5 which show the distributions
of the PSR sample including young PSRs. In these regions also, number of 
the SNRs with high surface brightness values is large. 

In the last 7 years, the Galactic plane (especially
the southern hemisphere) and
particularly the Galactic central directions were observed at 1400 MHz and 
a large number of new PSRs were found (Johnston et al. 1995; Manchester 
et al. 1996;
Sandhu et al. 1997; Lyne et al. 1998, 2000; Camilo et al. 2001; Edwards 
\& Bailes 2001a,b; Manchester 2001, D'Amico et al. 2001; Manchester et al.
2002; Morris et al. 2002). As a result of these searches, today the number 
of the known PSRs with measured 1400 MHz flux is larger than the number of 
the known PSRs with measured 400 MHz flux. Because of this, we  
examine the PSRs observed at 1400 MHz.

In Figure 4, 634 PSRs with $\mid$b$\mid$ $<$ 5$^o$ are displayed. 
As seen from the
flux distribution with respect to the Galactic longitude, many PSRs with
small F$_{1400}$ values are located in Galactic arms and in the Galactic
central directions. From the figure it is seen that, in the 280$^o$ $<$ l 
$<$ 340$^o$ part of the region which was searched with the highest 
sensitivity (F$_{1400}$ $<$ 0.2 mJy) more low-flux PSRs were found.
In the interval l = 0$^o$$\pm$20$^o$ the
number of PSRs with F$_{1400}$ $<$ 0.2 mJy is very small. The reason
for this is not the search being not so sensitive and precise, but the 
background
radiation being very strong. In Figure 5, F$_{1400}$ -- l diagram of 496
PSRs with $\mid$b$\mid$ $<$ 2$^o$ is displayed. When we compare
Figures 4 and 5, we see a larger decrease in the number of
the PSRs located in the Galactic central directions which have F$_{1400}$ 
$<$ 0.2 mJy. 

In Figure 6, $\mid$z$\mid$--l diagram of the PSRs with 
$\mid$l$\mid$$<$70$^o$, $\mid$b$\mid$ $<$ 2$^o$, DM $<$ 800 pc/cm$^3$ and 
F$_{1400}$ $<$ 0.5 mJy is represented. Since the electron density strongly 
depends on the longitude value, as the direction becomes far away
from the Galactic center direction, the PSRs (in such directions) which 
have the same DM value   
that the PSRs in the Galactic center direction have, are located at
larger distances. This leads to the possibility of the $\mid$z$\mid$
values to be larger for the same $\mid$b$\mid$ values. For the same DM  
value of two different PSRs, the smaller distance value belongs to the one 
which is closer to the Galactic center
direction. As seen from Figure 8, although the distances of the PSRs in 
the interval l = 0$^o$$\pm$20$^o$ are somewhat less, the average
$\mid$z$\mid$ value is larger and this shows that the average 
$\mid$b$\mid$ value of these PSRs is larger. This is also a result of the 
effect of the background radiation.  

In order to have the probability of observing the PSRs with roughly the
same flux values and the SNRs with roughly the same $\Sigma$ values to be
almost the same for the whole Galactic plane and in order not to reduce
the number of the objects too much, we will first consider only
the PSRs with F$_{1400}$ $\ge$ 0.2 mJy and the SNRs with $\Sigma$ (at 1 
GHz) $\ge$ 10$^{-21}$ Wm$^{-2}$Hz$^{-1}$ster$^{-1}$.

The character of the background radiation (see Figure 1) does not show the 
possibility of the influence of each of the OB associations separately on 
observations of PSRs and SNRs. Despite this fact, we have checked the 
possibility of influences of the OB associations given in the lists of 
Garmany \& Stencel (1992) and Melnik \& Efremov (1995). We did not find 
any significant contribution of any one of the OB associations to the 
Galactic background radiation.

\section{Discussion and Conclusions}
Observational data of PSRs (ATNF Pulsar Catalogue 2003; Guseinov et al. 
2003a) and
SNRs (Green 2001) show that even in the Galactic central
directions (l=0$^o$$\pm$10$^o$, $\mid$b$\mid$$<$2$^o$) all the SNRs with
$\Sigma$$>$10$^{-21}$ Wm$^{-2}$Hz$^{-1}$sr$^{-1}$ and the PSRs with  
F$_{1400}$$>$0.2 mJy are observable. Since the background radiation is
strongly dependent on the Galactic latitude, the SNRs with
$\Sigma$$>$1.5$\times$10$^{-22}$ Wm$^{-2}$Hz$^{-1}$sr$^{-1}$ in the same
longitude interval can be observed if $\mid$b$\mid$$>$4$^o$. Also, the   
SNRs in the interval 60$^o$ $<$ l $<$ 300$^o$ can easily be observed 
for all values of b if $\Sigma$$>$3$\times$10$^{-22}$ 
Wm$^{-2}$Hz$^{-1}$sr$^{-1}$. 

In the Galaxy, total number of the SNRs with
$\Sigma$$>$3$\times$10$^{-22}$
Wm$^{-2}$Hz$^{-1}$sr$^{-1}$ and d$<$3.2 kpc in the interval 60$^o$ $<$ l 
$<$ 300$^o$ is 33 (Guseinov et al. 2003b). It is seen from PSR-SNR 
associations that the ages of the SNRs which 
are genetically connected to PSRs do not exceed 3$\times$10$^4$ yr in 
general (Kaspi \& Helfand 2002). Since the SNRs in the regions we
examined have less surface brightness values on average, 
we can roughly say that the ages of these SNRs may
exceed 3$\times$10$^4$ yr but not greater than 5$\times$10$^4$ yr.
There are 23 SNRs with
$\Sigma$$>$10$^{-21}$ Wm$^{-2}$Hz$^{-1}$sr$^{-1}$ located at d$\le$3.2
kpc from the Sun, among which 14 of them are in the sector under 
consideration. If we assume that the ratio of 
the number of the SNRs with 3$\times$10$^{-22}$ $<$ $\Sigma$ $\le$
10$^{-21}$ Wm$^{-2}$Hz$^{-1}$sr$^{-1}$ to the number of the bright SNRs 
in the central region (l=0$^o$$\pm$60$^o$) is equal to the same ratio of 
the 
SNRs in the region 60$^o$ $<$ l $<$ 300$^o$, then we can use the ratio 
for the Galactic anticenter directions to find the number of dim SNRs in 
the Galactic central directions. In this case, the number of the SNRs with 
$\Sigma$$>$3$\times$10$^{-22}$ Wm$^{-2}$Hz$^{-1}$sr$^{-1}$ and ages 
$\le$5$\times$10$^4$ yr in the region up to 3.2 kpc from the Sun is 54. 
If we further assume that the radius of the Galaxy is 12 kpc and the 
average number density of the SNRs in the whole Galaxy is the same as the 
number density of the SNRs within 3 kpc around the Sun, then the number of 
the SNRs having ages less than
5$\times$10$^4$ yr must be about 800 in the Galaxy. (Since the 
distribution of SNRs in the Galaxy is not homogeneous and the distribution 
of their number density with respect to Galactic radius is not known well, 
we can not estimate number of the SNRs considerably better). From 
this result, the formation rate of SNRs turns out to be about one in 65 yr 
which is approximately the same as the SN explosion rate (van den Bergh \& 
Tammann 1991; Capellaro et al. 1999; Capellaro \& Turatto 2001). We can  
use the same approach to estimate the birth rate of PSRs. 

There are 48 PSRs with $\tau$$\le$10$^6$ yr located at d$\le$ 3.2 kpc
around the Sun (Guseinov et al. 2003a). If we assume the distribution of 
the PSRs in the Galaxy to be similar to the distribution of the SNRs given 
above, then the number of PSRs with $\tau$$\le$10$^6$ yr must be about 
710 in the Galaxy. If we further assume the beaming factor to be 
$\sim$0.35 
(Lyne \& Graham-Smith 1998), then number of the PSRs turns out to be 2030. 
We can estimate the total number of PSRs knowing that 75\% of the PSRs 
around the Sun have L$_{1400}$$>$3 mJy kpc$^2$ and using the luminosity 
function of Guseinov et al. (2003c): the number of PSRs with 
$\tau$$\le$10$^6$ yr must be $\sim$4.5$\times$10$^3$ in the Galaxy. Using 
this result the birth rate of PSRs is found to be one in 220 yr. But the 
PSRs with magnetic fields $>$10$^{13}$ G may pass the death belt in less 
than 10$^6$ yr. Taking this fact also into account, the birth rate of 
PSRs can roughly be assumed to be one in 200 yr. If we have used the 
luminosity function of Lorimer et al. (1993) or Allakhverdiev et al. 
(1997) instead of the luminosity function given by Guseinov et al. 
(2003c), then the birth rate of PSRs would be a bit larger.



\clearpage
\begin{figure}[t]
\vspace{3cm}
\includegraphics{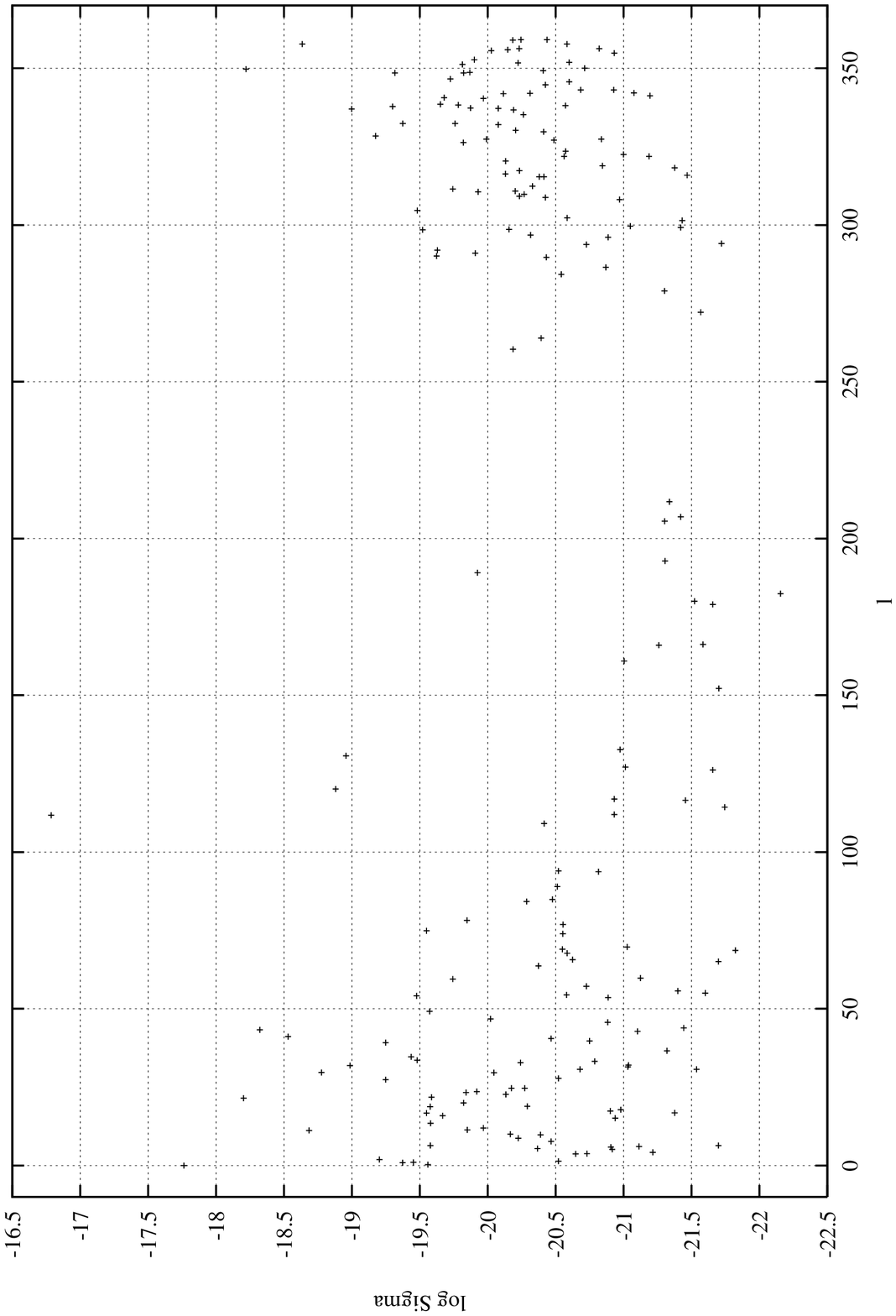}
\caption{Surface brightness versus longitude diagram for 201 SNRs with 
$\mid$b$\mid$$<$5$^o$.}
\end{figure}

\clearpage
\begin{figure}[t]
\vspace{3cm}
\includegraphics{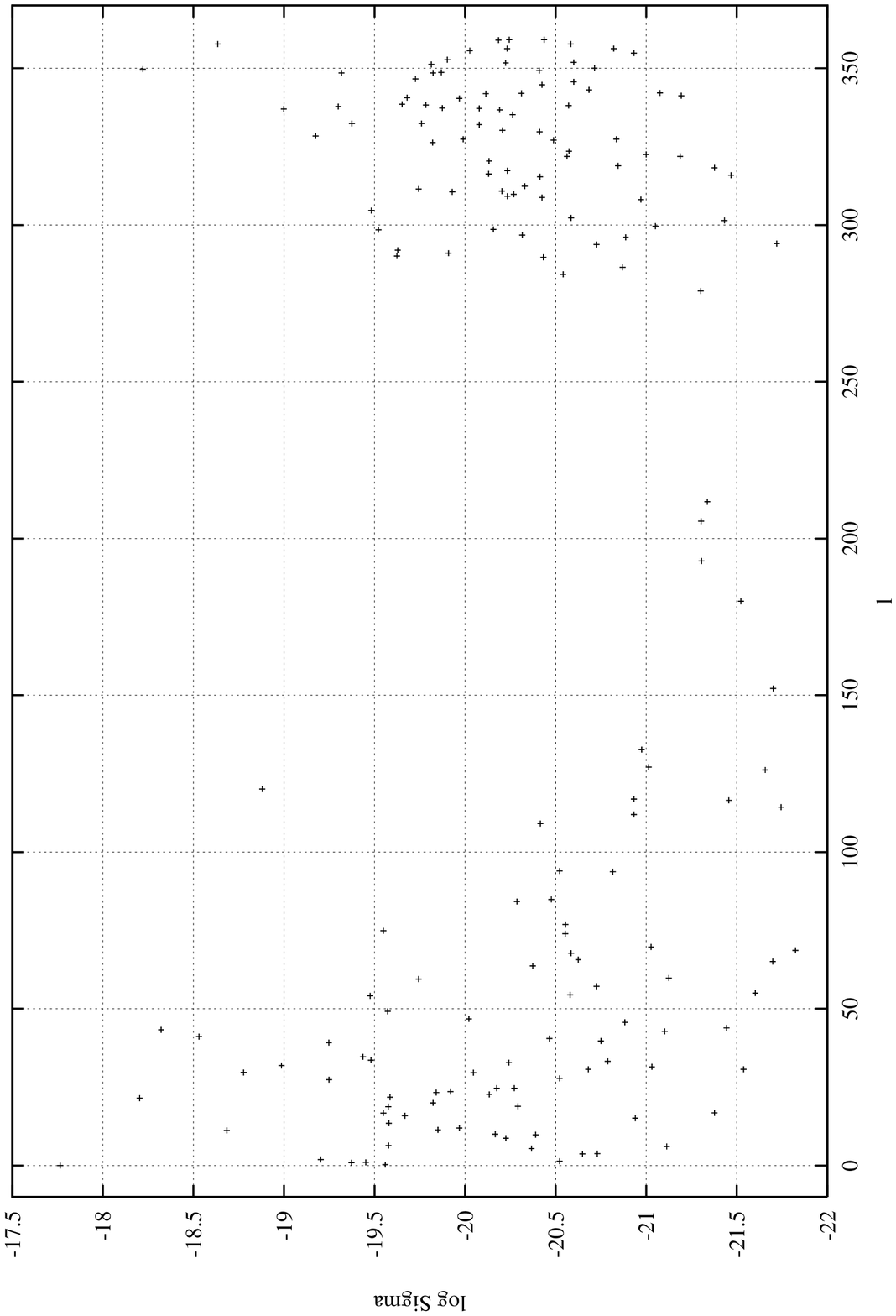}
\caption{Surface brightness versus longitude diagram for 171 SNRs with
$\mid$b$\mid$$<$2$^o$.}
\end{figure} 

\clearpage
\begin{figure}[t]
\vspace{3cm}
\includegraphics{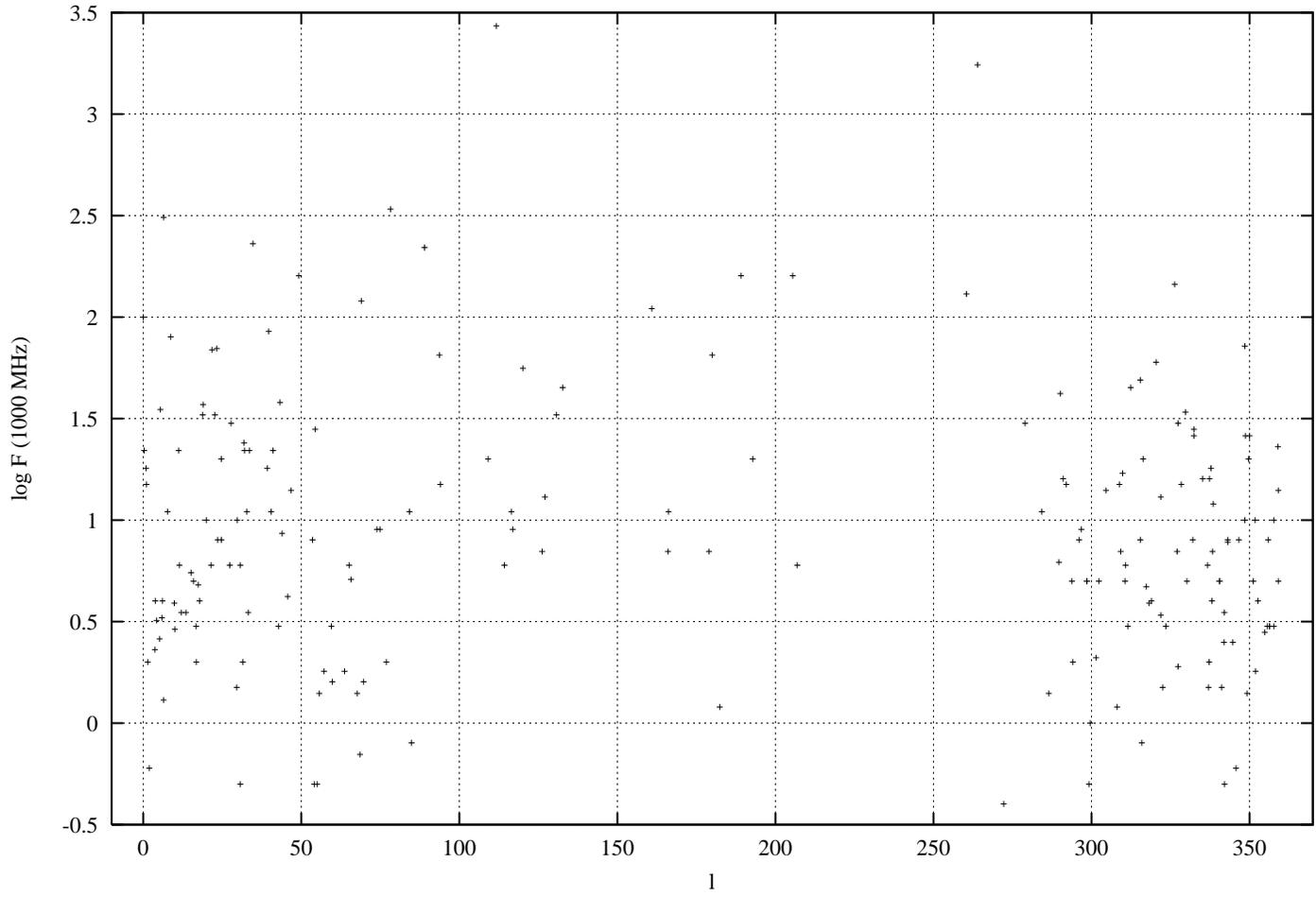}
\caption{F$_{1000}$ versus longitude diagram for 197 SNRs with 
$\mid$b$\mid$$<$5$^o$.}
\end{figure}

\clearpage
\begin{figure}[t]
\vspace{3cm}
\includegraphics{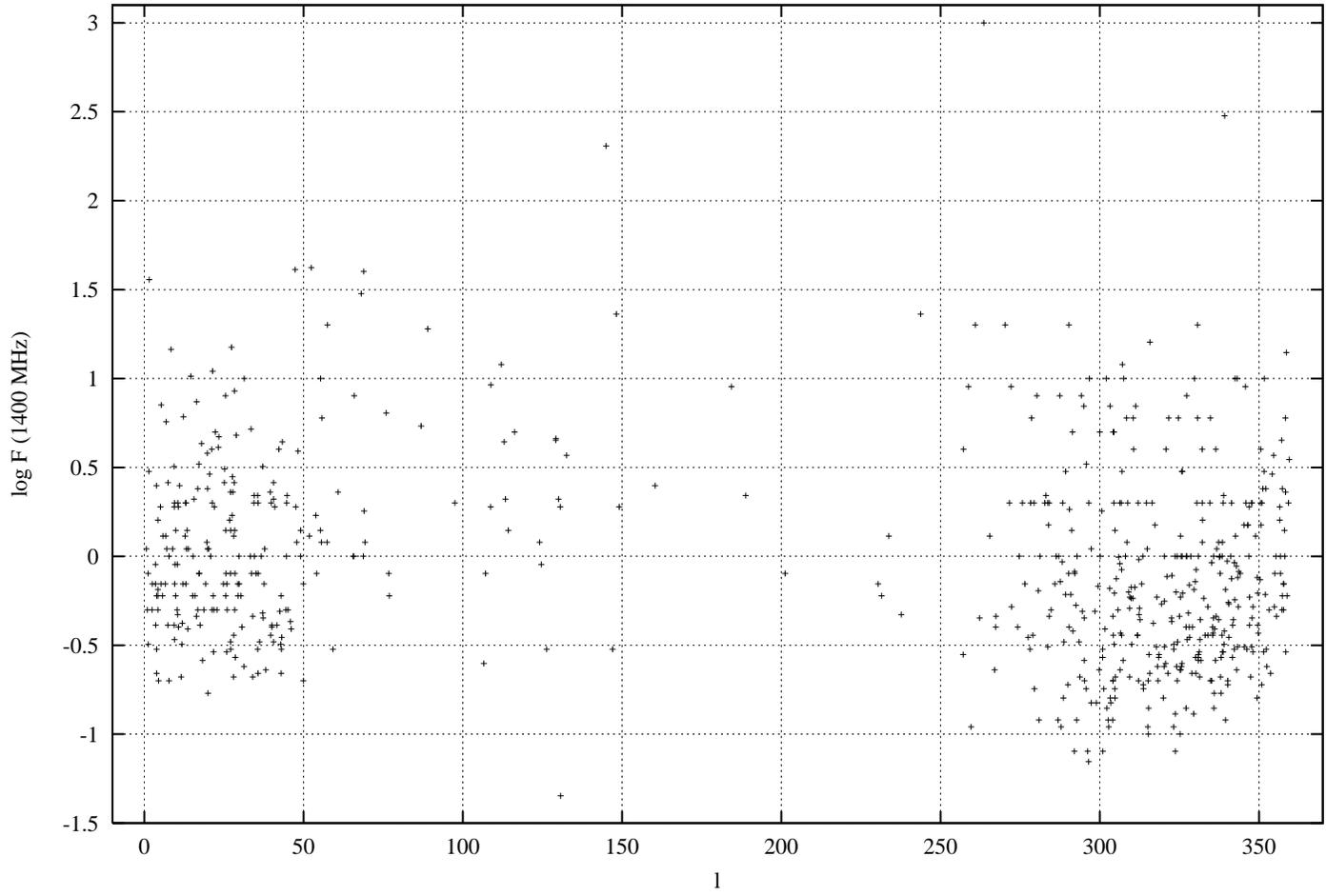}
\caption{F$_{1400}$ versus longitude diagram for 634 PSRs with 
$\mid$b$\mid$$<$5$^o$.}
\end{figure}

\clearpage
\begin{figure}[t]
\vspace{3cm}
\includegraphics{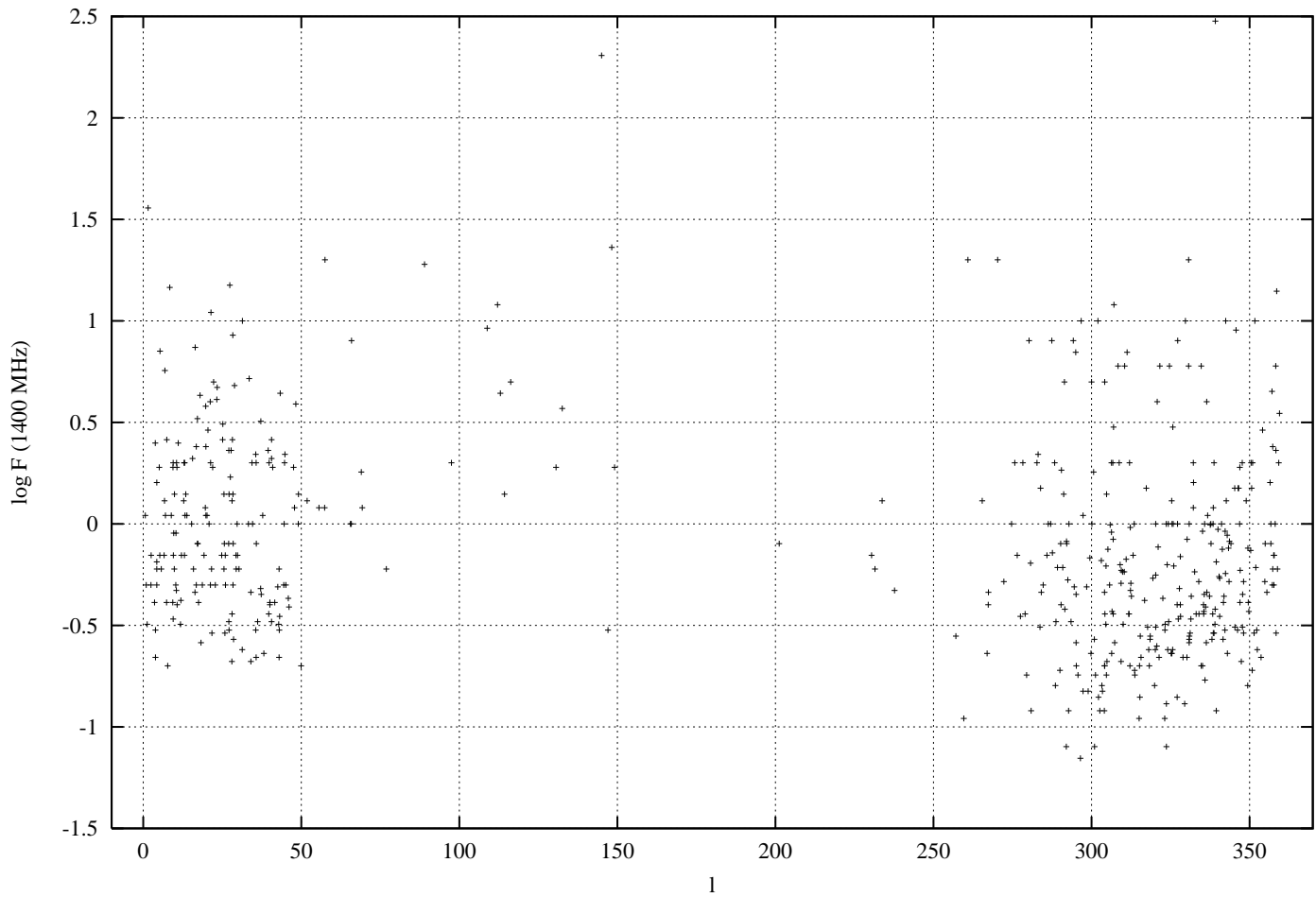}
\caption{F$_{1400}$ versus longitude diagram for 496 PSRs with         
$\mid$b$\mid$$<$2$^o$.}
\end{figure}

\clearpage
\begin{figure}[t]
\vspace{3cm}
\includegraphics{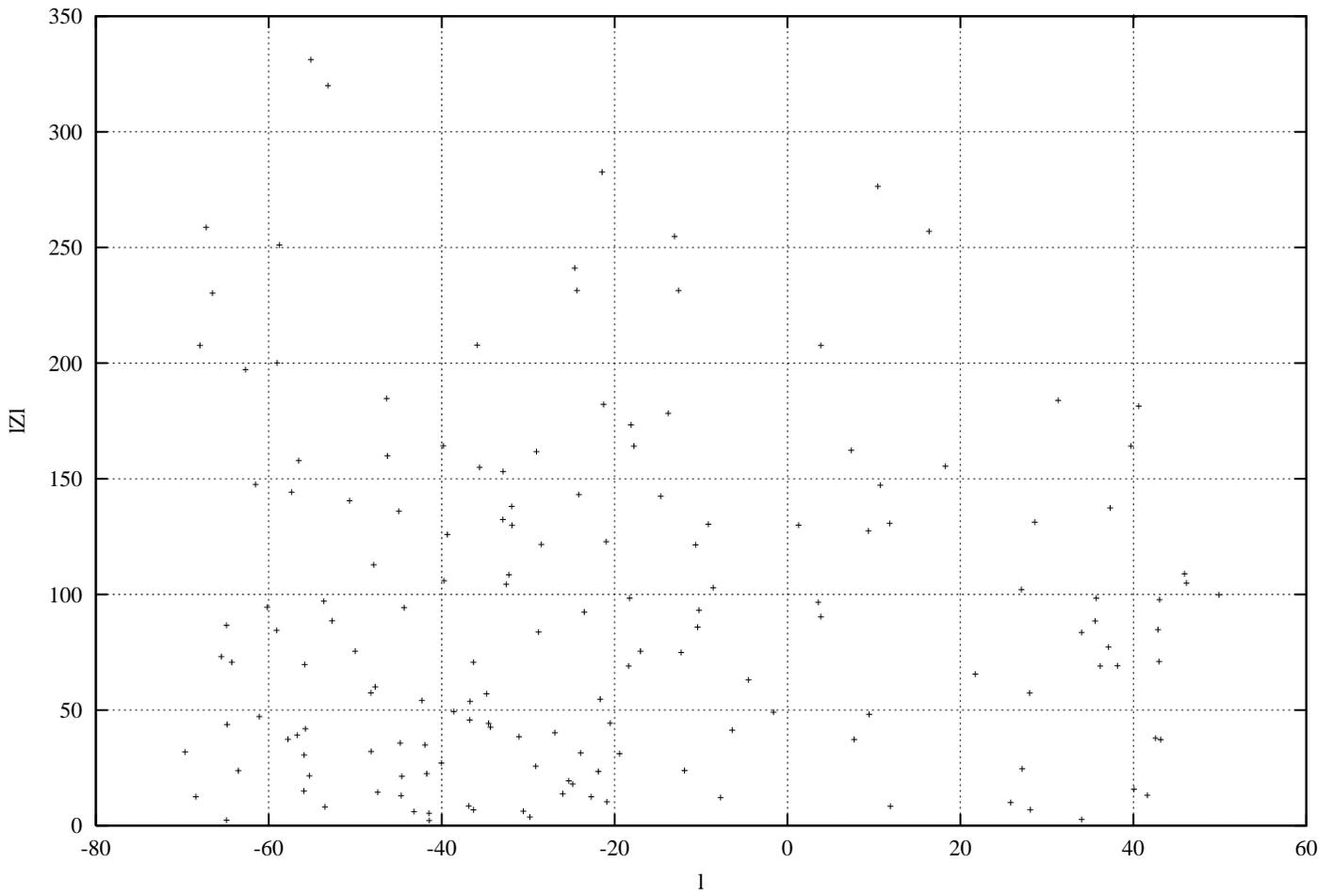}
\caption{$\mid$z$\mid$ versus longitude diagram of PSRs with 
$\mid$b$\mid$$<$2$^o$, F$_{1400}$$<$0.5 mJy, -70$^o$$<$l$<$55$^o$ and 
DM$<$800 pc/cm$^3$.} 
\end{figure}

\end{document}